\documentclass[a4paper,11pt]{article}
\pdfoutput=1 % if your are submitting a pdflatex (i.e. if you have
\usepackage{jheppub} % for details on the use of the package, please
\usepackage{amsmath}
\usepackage[T1]{fontenc} % if 
\def\bal#1\eal{\begin{align}#1\end{align}}
\def\alp[#1]{\begin{align}#1\end{align}}

\def\secnum[#1]{\texorpdfstring{$#1$}{TEXT}}

\def\secnuml#1\secnumr{\texorpdfstring{$#1$}{TEXT}}
\newcommand\beq{\begin{equation}}
\newcommand\eeq{\end{equation}}
\newcommand\bali{\begin{aligned}}
\newcommand\eali{\end{aligned}}
\def\eqa{\begin{eqnarray}}
\def\eqae{\end{eqnarray}}
\def\eq{\begin{equation}}
\def\eqe{\end{equation}}
\def\be{\begin{equation}}
\def\ee{\end{equation}}
\def\bea{\begin{eqnarray}}
\def\eea{\end{eqnarray}}
\def\ba{\begin{array}}
\def\ea{\end{array}}
\def\bd{\begin{displaymath}}
\def\ed{\end{displaymath}}

\def\>{\rangle}
\def\<{\langle}

\title{Causality Criteria for Island Models}

\author[a,b]{Feiyu Deng}

\affiliation[a]{Institute of Theoretical Physics, Chinese Academy of Sciences, Beijing 100190, China}
\affiliation[b]{School of Physical Sciences, University of Chinese Academy of Sciences, \\ Beijing 100049, China }
\emailAdd{dengfy@itp.ac.cn}

\abstract{
Island models provide a compelling resolution of the black hole information
paradox, but they also raise persistent questions regarding causal consistency in
effective descriptions.
In particular, effective theories appearing in double holography can exhibit
apparent violations of micro-causality, even though the underlying bulk dynamics
remains local and causal.
The aim of this work is to identify the physical origin of this phenomenon and to
clarify which structural features control causal consistency in island models.

We show that the apparent non-causality observed in double holography is neither
intrinsic to island physics nor a consequence of nonlocal operator
reconstruction.
Rather, it reflects a mismatch between the effective assignment of spacetime
separation and the causal accessibility determined by bulk dynamics, a feature
that can already be discerned in earlier analyses.
Nonlocal reconstruction instead captures quantum error correction within a
restricted code subspace and does not introduce independent propagation
channels.

Motivated by this perspective, we formulate a structural criterion for
micro-causality in effective descriptions of island models.
The criterion consists of three conditions: the absence of independent
propagation channels beyond those present in the bulk theory, a local
bulk-supported operator dictionary for effective operators, and a faithful
matching between effective spacelike separation and dynamically accessible bulk
causal curves.
When these conditions are satisfied, effective micro-causality follows directly
from bulk micro-causality.
We apply the criterion to brane world realizations of island models, including the
defect-extremal-surface construction, and show that they preserve causal
consistency, in contrast to double holography.
We further demonstrate that the criterion is robust under time-dependent
processes such as island formation and evaporation.
}

\begin{document}
\maketitle
\flushbottom
\section{Introduction}
\label{sec:introduction}
%%%%%%%%%%%%%%%%%%%%%%%%%%%%%%%%%%%%%%%%%%%%%%%%%%%%%%%%%%%%%%%%%%%%%%%%%%%%%%

Understanding how spacetime causality emerges from quantum degrees of freedom is a
central problem in quantum gravity.
In semiclassical gravity, causal structure is encoded geometrically through the
light cone of a smooth spacetime metric.
In holographic systems, by contrast, spacetime itself is an emergent notion,
arising from an underlying quantum theory whose organization is highly nonlocal.
Clarifying when and how effective descriptions inherit causal consistency from an
underlying bulk theory is therefore both conceptually fundamental and practically
important.

Island models have recently provided a powerful framework for addressing the
black hole information paradox
\cite{Almheiri:2019hni,Penington:2019npb,Almheiri:2019psy}.
By incorporating quantum extremal surfaces into the computation of entanglement
entropy, these models reproduce the Page curve in a wide class of gravitational
settings.
At the same time, island constructions place significant strain on conventional
notions of locality: degrees of freedom associated with the black hole interior
can be encoded in radiation degrees of freedom at infinity.
This feature has led to recurring concerns regarding the causal consistency of
effective island descriptions, especially when the effective theory does not
have explicit access to the bulk radial direction.
For recent developments on island physics, see
\cite{Antonini:2025sur,Yu:2025euq,Dey:2025dlj,Espindola:2025ons,Almheiri:2025ugo,
Padua-Arguelles:2025koj,Basu:2025sqk,Jiang:2025hao,Yu:2025tid,Basu:2024xjq,
Wen:2024uwr,Fumagalli:2024msi,Hao:2024nhd,Matsuo:2024ypr,Antonini:2024bbm,
Dey:2025qms,Santos:2024cvx,Liu:2024cmv,Lin:2023hzs,Jalan:2024cby,Jiang:2024xnd,Chang:2023gkt,Ahn:2021chg,Jeong:2023lkc}.

A particularly sharp arena in which these issues arise is double holography
\cite{Almheiri:2019psy}.
In this framework, an effective lower-dimensional gravitational description is
itself dual to a higher-dimensional bulk theory.
While the higher-dimensional bulk theory is manifestly local and causal, the
effective boundary description can exhibit apparent violations of
micro-causality: operators that are spacelike separated according to the
effective geometry may fail to commute.
This phenomenon has been analyzed in detail in previous work, notably in
\cite{Omiya:2021olc}, and provides a controlled setting in which effective
non-causality appears without any fundamental breakdown of bulk locality.

A central message of this paper is that the phrase ``double holography'' is commonly used to denote two structurally distinct constructions, which must be carefully
distinguished.
In what we will call \emph{bulk-first double holography}, island operators are
fundamentally bulk-supported: they admit representatives localized at finite bulk
points, and the effective description is derived from the bulk.
In this case, the bulk theory remains fully local and causal, and apparent
violations of micro-causality arise from a mismatch between effective spacetime
separation and bulk causal accessibility.
By contrast, in \emph{boundary-first double holography}, the Planck brane is
treated as an independent asymptotic boundary endowed with its own autonomous
operator algebra.
In this formulation, the minimal operator-level control needed to inherit bulk
micro-causality is absent from the outset.
Failing to distinguish these two notions obscures the origin of effective
non-causality and has led to confusion in the literature.

Another recurring source of misunderstanding concerns the role of nonlocal
operator reconstruction.
Island or brane operators often admit representations in terms of bath degrees of
freedom within a restricted code subspace, reflecting the quantum error-correcting
structure of holography
\cite{Almheiri:2014lwa,Pastawski:2015qua}.
Such reconstruction, however, is a statement about encoding rather than physical
signal propagation.
It does not introduce new dynamical channels through which information can be
transmitted.
Likewise, the bulk dynamics itself remains local and causal in all semiclassical
island constructions of interest.
Neither nonlocal reconstruction nor bulk dynamics alone is responsible for
effective non-causality.

The central thesis of this work is that causal consistency in island models is
controlled by the interplay of three logically independent structural ingredients:
(i) the availability of physical propagation channels,
(ii) the existence of a bulk-supported definition of effective operators,
and (iii) the interpretation of spacetime separation in the effective
description.
Conflating these ingredients obscures the mechanism by which effective
micro-causality can fail.

We make this statement precise by formulating a structural criterion for
micro-causality in effective descriptions of island models.
The criterion consists of three conditions:
\emph{localization}, requiring the absence of independent propagation channels
beyond those present in the bulk theory;
a \emph{bulk-supported operator dictionary}, ensuring that effective operators
admit local bulk representatives;
and \emph{matching}, requiring that effective spacelike separation faithfully
exclude all dynamically accessible bulk causal curves.
When these conditions are satisfied, effective micro-causality follows directly
from bulk micro-causality.

We apply this criterion to several classes of island constructions.
In bulk-first double holography, the localization and dictionary conditions are
satisfied, but the matching condition fails: effective spacelike separation does
not faithfully encode bulk causal accessibility, leading to apparent violations
of micro-causality.
In boundary-first double holography, the dictionary condition already fails, and
effective causality is not expected to follow from bulk locality.
By contrast, brane world realizations of island models, including the
defect-extremal-surface construction, satisfy all three conditions.
In these models, quantum field theory degrees of freedom are genuinely localized
on a codimension-one hypersurface and do not generate autonomous bulk light cones.
As a result, effective and bulk notions of spacetime separation coincide, and
micro-causality is preserved.

Finally, we show that the criterion is robust under time evolution.
Island formation, motion, and evaporation are intrinsically dynamical processes,
but time dependence does not introduce new sources of non-causality provided the
structural conditions are maintained locally in time.
This establishes the causal consistency of time-dependent island models realized
as brane world constructions.

Taken together, our results provide a structural characterization of
micro-causality in island models. In particular, the criterion we propose is not merely sufficient, but also
necessary in the sense that any violation of effective micro-causality must
originate from the failure of at least one of its components.
They clarify why bulk-first double holography exhibits apparent non-causality,
why boundary-first formulations fall outside the scope of bulk causal control,
why brane world constructions remain causal, and why nonlocal reconstruction
poses no threat to causal consistency.
More broadly, the analysis illustrates how causal structure can emerge—or fail to
emerge—in effective gravitational descriptions of quantum systems.

%%%%%%%%%%%%%%%%%%%%%%%%%%%%%%%%%%%%%%%%%%%%%%%%%%%%%%%%%%%%%%%%%%%%%%%%%%%%%%
\section{Review of double holography}
\label{sec:review-dh}
%%%%%%%%%%%%%%%%%%%%%%%%%%%%%%%%%%%%%%%%%%%%%%%%%%%%%%%%%%%%%%%%%%%%%%%%%%%%%%

In this section we review the framework commonly referred to as
\emph{double holography}, emphasizing a distinction that will be crucial for our
subsequent analysis of effective causality.
Rather than treating double holography as a single, monolithic construction, we
will carefully separate two structurally different implementations that are
often discussed under the same name.
As we will see, this distinction is immaterial for many entanglement-based
applications, but it is decisive for questions of operator locality and
micro-causality.

\vspace{0.5em}
\noindent
\textbf{General setup.}
Double holography refers broadly to a situation in which a single physical system
admits multiple interrelated descriptions connected by holographic duality
applied at different levels.
The paradigmatic example involves three layers:
a lower-dimensional quantum mechanical description,
an intermediate effective boundary description with dynamical gravity,
and a higher-dimensional bulk gravitational description.
The coexistence of these descriptions relies on a sequence of nontrivial
assumptions about holographic duality and the organization of degrees of freedom
\cite{Almheiri:2019psy}.

We begin with a semiclassical system consisting of two-dimensional gravity
coupled to a two-dimensional conformal field theory, which is in turn coupled at
its boundary to a non-gravitational CFT bath defined on a fixed flat background.
The gravitational sector is treated semiclassically: the metric is dynamical, but
quantum fluctuations of geometry are suppressed by a large central charge.
Throughout this work, we restrict attention to regimes in which the bulk
description admits a local notion of micro-causality.

Applying the AdS/CFT correspondence to the non-gravitational CFT degrees of
freedom, the combined system admits a dual description in terms of a
three-dimensional asymptotically AdS spacetime.
In this bulk description, the original two-dimensional gravity theory is not
eliminated; instead, it is encoded through a dynamical boundary condition imposed
on a codimension-one hypersurface, commonly referred to as the Planck brane.
The asymptotic boundary of AdS$_3$ hosts the bath CFT and is endowed with Dirichlet
boundary conditions, while the Planck brane carries induced gravity and
intersects the asymptotic boundary along a one-dimensional locus
\cite{Karch:2000gx,Takayanagi:2011zk}.

\begin{figure}[h] \centering \includegraphics[width=15cm,height=5cm]{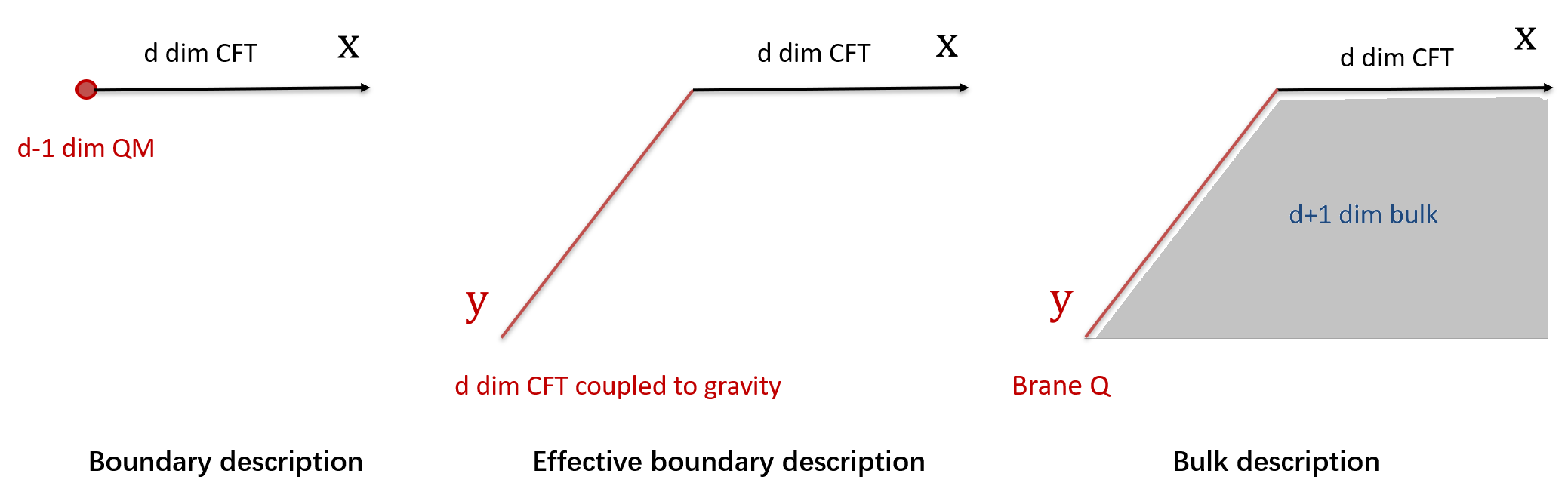} \caption{Three interrelated descriptions in double holography: the one-dimensional boundary description, the two-dimensional effective boundary description with gravity, and the three-dimensional bulk description with a Planck brane.} \label{dh} \end{figure}

A further essential ingredient is the assumption that the two-dimensional gravity
theory coupled to the CFT itself admits a dual description in terms of a
one-dimensional quantum mechanical system.
Under this assumption, the same physical setup can be described in three
languages:
\begin{itemize}
\item a one-dimensional boundary description, given by quantum mechanics;
\item a two-dimensional effective boundary description, given by a BCFT coupled
to two-dimensional gravity;
\item a three-dimensional bulk description, given by AdS$_3$ gravity with a
Planck brane.
\end{itemize}
These three descriptions are schematically illustrated in Fig.~\ref{dh}.
What is usually called \emph{double holography} refers to the simultaneous
existence of these descriptions.

\vspace{0.8em}
\noindent
\textbf{Two notions of double holography.}
At this point, it is essential to distinguish two inequivalent structural
interpretations of the above setup.
In the following, we will refer to these as \emph{boundary-first double
holography} and \emph{bulk-first double holography}.
Although both arise from the same geometric construction, they differ in how the
Planck brane is treated at the operator level, and this difference has direct
consequences for causality.

\vspace{0.5em}
\noindent
\textbf{Boundary-first double holography.}
In the boundary-first interpretation \cite{Almheiri:2019psy}, the Planck brane is treated on the same
conceptual footing as an asymptotic boundary.
The CFT living on the brane is regarded as an autonomous boundary theory, endowed
with its own intrinsic operator algebra.
One then applies AdS/CFT to the brane degrees of freedom in the same sense as for
the bath CFT, effectively promoting the brane to a second boundary of the bulk
spacetime.

In this viewpoint, operators on the brane are not required to admit a
bulk-supported realization at a finite bulk locus.
Instead, they are treated as independent boundary operators whose algebraic
relations are not fixed solely by bulk locality.
While this interpretation is natural from the perspective of entanglement
entropy and state counting, it obscures the relation between effective operator
commutators and bulk micro-causality.
As we will discuss later, this boundary-first viewpoint lies outside the regime
in which effective causality can be directly inferred from bulk locality.

\vspace{0.5em}
\noindent
\textbf{Bulk-first double holography.}
In contrast, the bulk-first interpretation treats the Planck brane as a finite
bulk locus rather than as an autonomous boundary \cite{Neuenfeld:2021wbl,Omiya:2021olc}.
The brane supports dynamical degrees of freedom, but these degrees of freedom are
not endowed with an independent boundary operator algebra.
Instead, operators associated with the brane are understood to admit
bulk-supported representatives localized at finite bulk points on the brane
worldvolume.

This is the interpretation implicitly adopted in analyses of causality in double
holography, including the work of Wei and collaborators \cite{Omiya:2021olc}.
In this bulk-first viewpoint, the bulk theory remains manifestly local and
causal, and the Planck brane does not introduce new asymptotic regions or
independent causal structures.
Any apparent violation of causality in the effective boundary description must
therefore arise from how spacetime separation is defined after integrating out
the bulk radial direction, rather than from a breakdown of bulk micro-causality
or from an autonomous boundary operator algebra.

\vspace{0.8em}
\noindent
\textbf{Concrete realization.}
To make these distinctions explicit, we consider a standard geometric
realization.
The effective boundary description is taken to be a two-dimensional topological
gravity theory coupled to a CFT living on a fixed AdS$_2$ background, which is
itself coupled at its boundary to a flat CFT bath.
From the bulk perspective, this system is described by AdS$_3$ gravity with a
Planck brane whose intrinsic geometry is AdS$_2$, and on which the topological
gravity theory is localized \cite{Almheiri:2019yqk}.

In Poincar\'e coordinates, the AdS$_3$ metric takes the form
\begin{align}
d s^{2}_3
&=
\frac{l^2}{z^2}\left(-dt^2+dz^2+dx^2\right) ,
\end{align}
where $l$ is the AdS radius.
It is often convenient to rewrite this metric in coordinates adapted to an
AdS$_2$ slicing.
Introducing coordinates $(\rho,y)$, or equivalently an angular coordinate
$\theta$, the metric can be expressed as
\begin{align}
d s^{2}_3
&=
d \rho^{2}
+
l^2\cosh ^{2} \frac{\rho}{l}
\cdot
\frac{-d t^{2}+d y^{2}}{y^{2}}
\nonumber\\
&=
\frac{1}{\cos^2\theta}
\left(
d\theta^2
+
l^2
\cdot
\frac{-d t^{2}+d y^{2}}{y^{2}}
\right) .
\label{2dmet}
\end{align}
These coordinate systems are related to the Poincar\'e coordinates by
\begin{align}
z
&=
-y \cosh \frac{\rho}{l}
=
-y\cos\theta ,
\nonumber\\
x
&=
y \tanh \frac{\rho}{l}
=
y\sin\theta .
\label{zyxy}
\end{align}

The asymptotic boundary $\Sigma$ of AdS$_3$ is located at $z=0$ and is parametrized
by the coordinates $(x,t)$.
A codimension-one hypersurface $Q$ defined by fixing $\rho=\rho_0$, or
equivalently $\theta=\theta_0$, has induced metric
\begin{equation}
ds^2_Q
=
l^2\cosh^2\!\frac{\rho_0}{l}
\cdot
\frac{-dt^2+dy^2}{y^2} ,
\end{equation}
which is AdS$_2$.
In the bulk-first interpretation adopted throughout this work, this hypersurface
is treated as a finite bulk locus rather than as an asymptotic boundary.

An observer confined to the effective boundary description does not have access
to the bulk radial direction.
Consequently, spacetime separation and causal relations are defined intrinsically
within the two-dimensional effective geometry.
The relation between the coordinate $x$ on the asymptotic boundary $\Sigma$ and
the coordinate $y$ on the brane $Q$ is fixed by the transparent boundary condition
at their intersection, together with the matching of stress tensors on $\Sigma$
and $Q$.
As shown in \cite{Almheiri:2019yqk}, this leads to the simple
identification
\begin{equation}
x = y .
\label{xy}
\end{equation}
This identification plays a central role in the effective boundary description.
In the following sections, we will show that it is precisely the interpretation
of spacetime separation implied by this identification that leads to apparent
violations of micro-causality in bulk-first double holography.

%%%%%%%%%%%%%%%%%%%%%%%%%%%%%%%%%%%%%%%%%%%%%%%%%%%%%%%%%%%%%%%%%%%%%%%%%%%%%%
\section{Non-causality in bulk-first double holography}
\label{sec:noncausality-dh}
%%%%%%%%%%%%%%%%%%%%%%%%%%%%%%%%%%%%%%%%%%%%%%%%%%%%%%%%%%%%%%%%%%%%%%%%%%%%%%

In this section we analyze the origin of apparent non-causality in the effective
boundary description of \emph{bulk-first double holography}.
Throughout this section, the Planck brane is treated as a finite bulk locus rather
than as an autonomous asymptotic boundary.
Operators associated with the brane are assumed to admit bulk-supported
representatives at finite bulk locations, and the bulk theory itself is taken to
be manifestly local and causal.
The non-causality discussed below therefore does not originate from a breakdown
of bulk micro-causality or from an autonomous boundary operator algebra on the
brane, but from a mismatch between effective spacetime separation and dynamically
accessible bulk causal relations.

We consider a spatial interval located entirely in the bath region on the
asymptotic boundary $\Sigma$.
Its associated island is determined by quantum extremal surfaces that emanate
from the endpoints of the interval and terminate on the Planck brane $Q$.
For simplicity, we restrict attention to a fixed spacelike time slice, on which
entanglement wedges and causal wedges are unambiguously defined.

\begin{figure}[h]
  \centering
  \includegraphics[width=15cm,height=4.8cm]{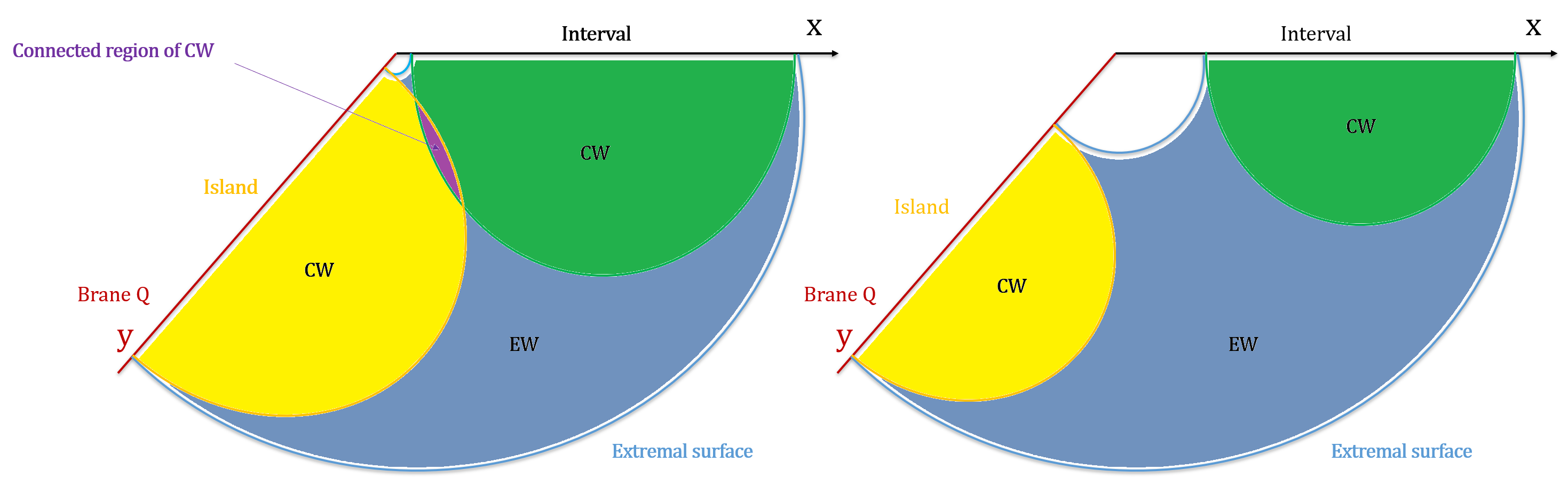}
  \caption{Causal wedges of a bath interval and its island can either intersect (left) or remain disjoint (right)
 in bulk-first double holography, despite the regions
  being disconnected in the effective boundary description.}
  \label{cw}
\end{figure}

A characteristic geometric feature of bulk-first double holography is that,
although the bath interval and its island are disconnected regions in the
effective boundary description, their associated causal wedges in the bulk need
not be disjoint.\footnote{
Here the phrase “causal wedge of the island” is used in the bulk-first sense.
It refers to the bulk causal wedge of the bulk region that is interpreted as the
island in the effective boundary description, rather than to a causal wedge
defined intrinsically by an autonomous boundary or defect theory.
In particular, the island is not treated as an independent boundary system with
its own causal structure; the causal wedge is defined entirely by the bulk light
cone and only subsequently interpreted in boundary terms.
}
Depending on the geometric configuration, the causal wedge of the bath interval
may intersect that of its island, or may remain separated, as illustrated in
Fig.~\ref{cw}.
Such intersections do not occur for pairs of regions both localized on $Q$ or
both localized on $\Sigma$.
They arise only in mixed configurations involving one region on the Planck brane
and one on the asymptotic boundary, and therefore reflect a genuinely mixed
bulk–boundary phenomenon \cite{Omiya:2021olc}.

\medskip
\noindent
\textbf{Clarifying remark.}
The intersection of causal wedges in Fig.~\ref{cw} should not be interpreted as
the appearance of an additional propagation channel.
In bulk-first double holography, operators associated with the island are bulk
operators rather than defect-localized degrees of freedom.
As a result, the bulk theory contains a single causal structure, and the bulk
dynamics remains entirely local and causal.
The significance of this observation will be formalized later as part of the
causality criterion.

From the perspective of an observer confined to the effective boundary
description, such bulk causal connections are not directly visible.
The effective theory assigns spacetime separation using only intrinsic boundary
coordinates and therefore may declare two operator insertions to be spacelike
separated even when a bulk causal curve exists between their bulk-supported
representatives.
This mismatch between effective spacetime separation and bulk causal
accessibility is the source of the apparent non-causality.
We now make this statement precise at the level of operator commutators.

\begin{figure}[h]
  \centering
  \includegraphics[width=12cm,height=7cm]{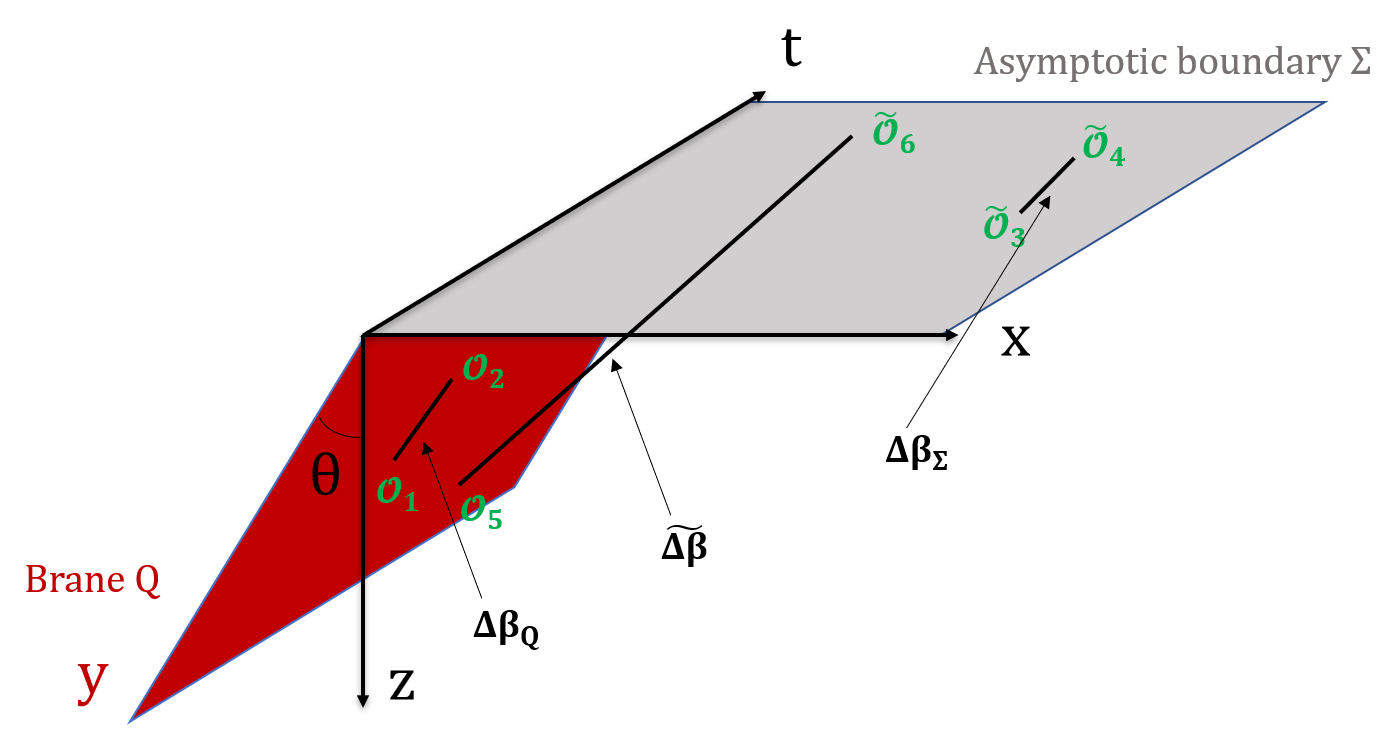}
  \caption{Distinct notions of spacetime separation for operators localized on
  the Planck brane $Q$ and the asymptotic boundary $\Sigma$, viewed from the bulk
  perspective.}
  \label{ss1}
\end{figure}

Micro-causality requires that local operators commute whenever they are spacelike
separated.
In bulk-first double holography, however, several inequivalent notions of
spacetime separation coexist.
From the bulk perspective, one may define the following invariant separations
between two points $p$ and $p'$:
\begin{align}
\Delta \beta_Q(p,p')
&=
-(t-t')^2+(y-y')^2 ,
\label{qq}
\\
\Delta \beta_\Sigma(p,p')
&=
-(t-t')^2+(x-x')^2 ,
\label{ss}
\\
\widetilde{\Delta \beta}(p,p')
&=
-(t-t')^2+(x-x')^2+(z-z')^2 ,
\label{busq}
\end{align}
where $\Delta \beta_Q$ and $\Delta \beta_\Sigma$ denote the intrinsic separations
along the Planck brane $Q$ and the asymptotic boundary $\Sigma$, respectively,
while $\widetilde{\Delta \beta}$ is the full bulk invariant separation.
These notions are illustrated schematically in Fig.~\ref{ss1}.

\begin{figure}[h]
  \centering
  \includegraphics[width=12cm,height=8cm]{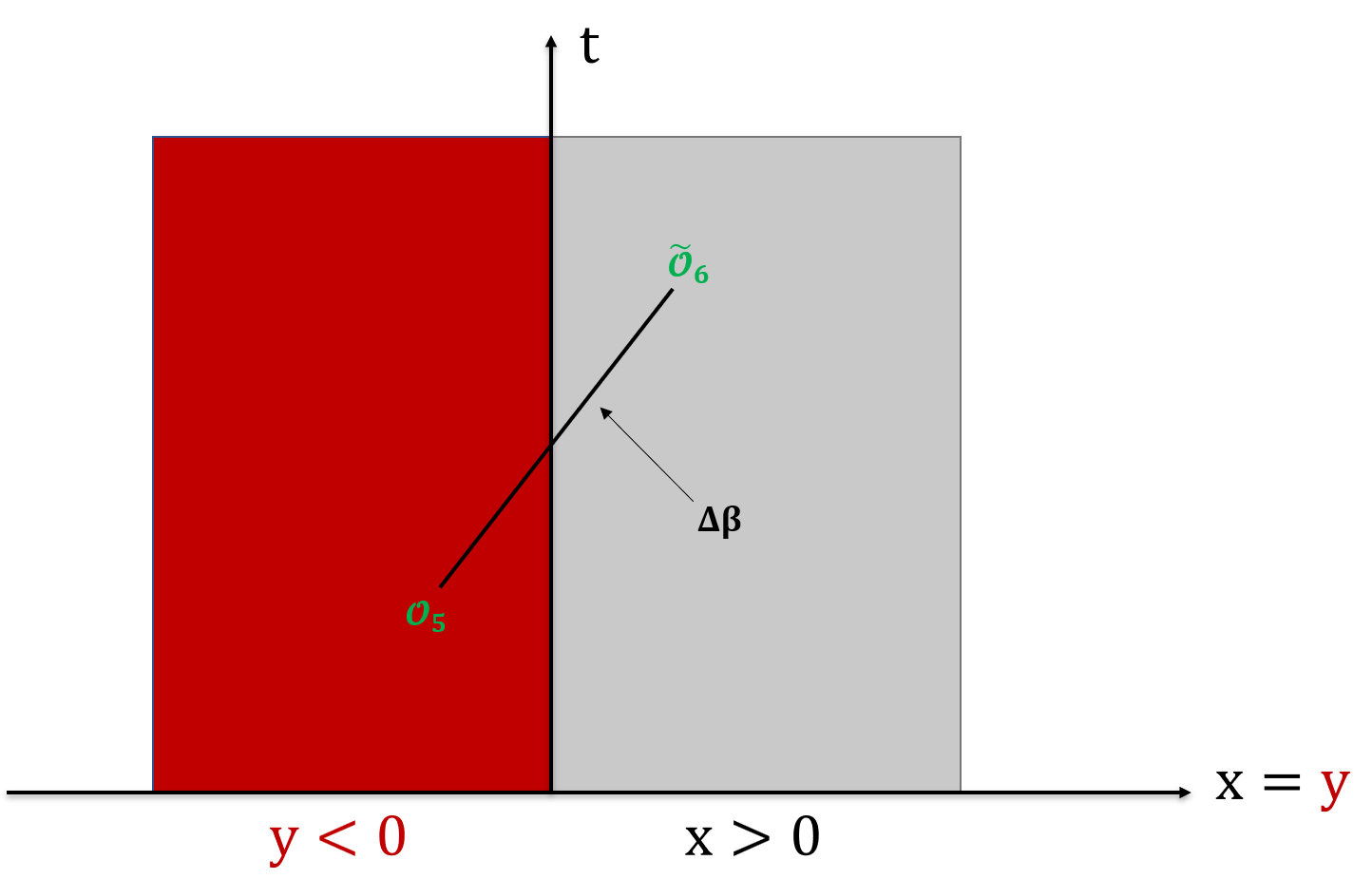}
  \caption{Effective spacetime separation between an operator on $Q$ and an
  operator on $\Sigma$, defined intrinsically in the effective boundary
  description.}
  \label{ss2}
\end{figure}

An observer confined to the effective boundary description does not have access
to the bulk radial direction.
Consequently, spacetime separation must be assigned intrinsically in terms of the
boundary coordinates $(t,x)$ on $\Sigma$ and $(t,y)$ on $Q$.
For pairs of operators both localized on $Q$ or both localized on $\Sigma$, the
effective separation coincides with $\Delta \beta_Q$ or $\Delta \beta_\Sigma$,
respectively.
For mixed configurations involving one operator on $Q$ and one on $\Sigma$,
however, the effective description assigns the separation
\begin{align}
\Delta \beta(p,p')
&=
-(t-t')^2+(x-y')^2
\nonumber\\
&=
-(t-t')^2+(x-x')^2
\nonumber\\
&=
-(t-t')^2+(y-y')^2 ,
\label{busq'}
\end{align}
as illustrated in Fig.~\ref{ss2}.
Effective spacelike separation is therefore characterized by $\Delta \beta>0$.

To assess micro-causality, one must evaluate commutators of operators inserted at
points that are declared spacelike separated by the effective geometry.

Operators localized on the asymptotic boundary $\Sigma$ are defined by the
standard extrapolate dictionary of AdS/CFT,
\begin{align}
\left\langle
\widetilde{\mathcal O}(t,x)
\widetilde{\mathcal O}(t',x')
\right\rangle_{\Sigma}
=
\lim_{z\to 0}
z^{-2\Delta}
\left\langle
\phi(z,t,x)\phi(z,t',x')
\right\rangle_{\text{bulk}} ,
\label{ex}
\end{align}
where $\widetilde{\mathcal O}$ is a primary operator of scaling dimension $\Delta$
and $\phi$ is the corresponding bulk field.
Bulk micro-causality ensures that operators on $\Sigma$ commute whenever
$\Delta \beta_\Sigma>0$.

Operators associated with the Planck brane $Q$ are assumed to admit
bulk-supported representatives localized at finite bulk points on the brane
worldvolume.
Under this assumption, correlators of operators on $Q$ reduce to bulk correlators
evaluated at fixed bulk locations, and operators on $Q$ commute whenever
$\Delta \beta_Q>0$ \cite{Neuenfeld:2021wbl,Omiya:2021olc}.

For mixed correlators involving one operator on $Q$ and one on $\Sigma$, the
relevant bulk-supported representatives are located at finite bulk points
$X_Q(p)$ and at the asymptotic boundary $X_\Sigma(p')$, respectively.
Bulk micro-causality implies that the bulk commutator
$\big[\phi(X),\phi(X')\big]$ vanishes whenever the bulk invariant separation
$\widetilde{\Delta \beta}(X,X')$ is positive.

Consider an operator $\mathcal O_5$ localized on $Q$ and an operator
$\widetilde{\mathcal O}_6$ localized on $\Sigma$, as shown in Fig.~\ref{ss2}.
In the effective description their coordinates are $(t,y)$ and $(t',x')$,
respectively, while in the bulk they correspond to the points
\begin{align}
\mathcal O_5 &: (t,\,y\sin\theta,\,-y\cos\theta),\\
\widetilde{\mathcal O}_6 &: (t',\,x',\,0).
\end{align}
The bulk invariant separation between these points is
\begin{align}
\widetilde{\Delta \beta}
&=
-(t-t')^2+(y\sin\theta-x')^2+y^2\cos^2\theta
\nonumber\\
&=
-(t-t')^2+(y-x')^2+2x'y(1-\sin\theta)
\nonumber\\
&=
\Delta \beta + 2x'y(1-\sin\theta) .
\end{align}

The condition for vanishing of the bulk commutator,
$\widetilde{\Delta \beta}>0$, therefore does not coincide with the effective
spacelike condition $\Delta \beta>0$.
Since $y<0$ and $\sin\theta<1$, it is possible for $\Delta \beta$ to be positive
while $\widetilde{\Delta \beta}$ is negative.
In such cases, two operators that are spacelike separated according to the
effective boundary geometry are nevertheless timelike related in the bulk, and
their commutator does not vanish.

We conclude that the apparent non-causality of the effective boundary description
in bulk-first double holography originates from a failure of effective spacetime
separation \emph{within the effective boundary description} to faithfully encode dynamically accessible bulk causal relations.
The bulk theory itself remains local and causal, and operators admit
bulk-supported representatives at finite bulk locations.
The non-causality therefore reflects a breakdown of matching between effective
and bulk notions of spacetime separation, rather than any failure of bulk
micro-causality or of the operator dictionary.
This observation motivates the formulation of a general criterion for
micro-causality in island models, to which we now turn.

%%%%%%%%%%%%%%%%%%%%%%%%%%%%%%%%%%%%%%%%%%%%%%%%%%%%%%%%%%%%%%%%%%%%%%%%%%%%%%
%%%%%%%%%%%%%%%%%%%%%%%%%%%%%%%%%%%%%%%%%%%%%%%%%%%%%%%%%%%%%%%%%%%%%%%%%%%%%%
\section{Nonlocal reconstruction versus causal propagation}
\label{sec:reconstruction-vs-propagation}
%%%%%%%%%%%%%%%%%%%%%%%%%%%%%%%%%%%%%%%%%%%%%%%%%%%%%%%%%%%%%%%%%%%%%%%%%%%%%%

In island models admitting a semiclassical bulk description, operators localized
on islands or branes may often be reconstructed in terms of degrees of freedom
in the bath.
Such reconstructions are typically spatially nonlocal when viewed from the
effective boundary theory.
This feature is a generic consequence of holographic encoding and quantum error
correction, and does not by itself imply any modification of causal dynamics
\cite{Almheiri:2014lwa,Pastawski:2015qua}.

Concretely, within a suitable code subspace $\mathcal H_{\rm code}$, one may
encounter relations of the form
\begin{equation}
\mathcal O_Q \,|\psi\rangle
=
\mathcal O^{\rm rec}_\Sigma \,|\psi\rangle,
\qquad
|\psi\rangle \in \mathcal H_{\rm code},
\label{eq:codesubspace-reconstruction}
\end{equation}
where $\mathcal O_Q$ denotes an operator localized on a codimension-one surface
$Q$, $\mathcal O^{\rm rec}_\Sigma$ is an operator constructed from degrees of
freedom on the asymptotic boundary $\Sigma$, and $\mathcal H_{\rm code}$ is a
restricted subspace of the full Hilbert space.
Such relations express the redundancy of the holographic encoding of bulk
information and are characteristic of quantum error-correcting codes.

Equation \eqref{eq:codesubspace-reconstruction} does not represent an operator
identity on the full Hilbert space.
It asserts equivalence only at the level of action on states belonging to
$\mathcal H_{\rm code}$.
Outside this subspace, the operators $\mathcal O_Q$ and
$\mathcal O^{\rm rec}_\Sigma$ generally act differently and need not share the
same algebraic properties.
In particular, code-subspace reconstruction places no constraint on operator
commutators evaluated as algebraic objects, nor does it modify the causal
relations dictated by the underlying dynamics.

Causal propagation, by contrast, is a dynamical notion.
It refers to the existence of physical channels through which excitations created
by an operator can influence other operators at later times.
In a local bulk quantum field theory, such channels are constrained by the bulk
light-cone structure, and micro-causality is encoded in the vanishing of
commutators at spacelike separation.
Whether two operator insertions can influence one another is therefore determined
by the bulk equations of motion and their causal structure, not by the
availability of alternative operator representations.

Nonlocal reconstruction does not introduce new propagation channels.
The existence of a reconstruction operator $\mathcal O^{\rm rec}_\Sigma$ that
reproduces the action of $\mathcal O_Q$ on $\mathcal H_{\rm code}$ does not allow
signals created by $\mathcal O_Q$ to propagate outside the bulk light cone.
Reconstruction is a statement about how information is encoded in the boundary
theory, not about how excitations propagate in spacetime.

This distinction becomes sharp at the level of commutators.
Consider a bath operator $\widetilde{\mathcal O}_\Sigma(p')$ localized at a point
$p'\in\Sigma$ and an operator $\mathcal O_Q(p)$ localized at a point $p\in Q$.
Even if $\mathcal O_Q$ admits a highly nonlocal reconstruction
$\mathcal O^{\rm rec}_\Sigma$ with extended support on $\Sigma$, the commutator
\begin{equation}
\bigl[\mathcal O_Q(p), \widetilde{\mathcal O}_\Sigma(p')\bigr]
\label{eq:recon-commutator}
\end{equation}
is governed by the causal relation between the corresponding bulk-supported
operators, provided that $\mathcal O_Q$ admits a local bulk representation and
does not generate independent propagation channels.
The reconstruction operator $\mathcal O^{\rm rec}_\Sigma$ does not constitute an
additional dynamical degree of freedom; it merely provides an alternative
representation of the same bulk excitation within a restricted subspace.

As a result, nonlocal reconstruction is fully compatible with strict
micro-causality.
Violations of effective micro-causality cannot arise from reconstruction alone,
but require the existence of a dynamically accessible bulk causal curve that is
not excluded by the effective notion of spacetime separation.
When the effective description correctly accounts for all available propagation
channels, nonlocal reconstruction of island or brane operators does not lead to
any conflict with causality.

The causality criterion formulated in the following section makes this separation
precise.
Its purpose is not to constrain nonlocal reconstruction, which is an intrinsic
feature of holographic encoding, but to identify the conditions under which
effective spacelike separation faithfully reflects the absence of any
dynamically accessible bulk causal curve.
Under these conditions, the nonlocality of reconstruction resides entirely in
the encoding of information and has no bearing on the causal structure of the
underlying dynamics.

\section{A sufficient criterion for micro-causality in island models}
\label{subsec:causality-criterion}
%%%%%%%%%%%%%%%%%%%%%%%%%%%%%%%%%%%%%%%%%%%%%%%%%%%%%%%%%%%%%%%%%%%%%%%%%%%%%%

The analysis of the previous section shows that the apparent violation of
micro-causality in bulk-first double holography does not originate from any
failure of bulk locality, nor from nonlocal operator reconstruction.
Rather, it arises because the effective boundary description assigns spacetime
separation in a way that does not faithfully reflect which bulk causal curves are
physically accessible.
Motivated by this observation, we now formulate a \emph{sufficient} criterion
under which an effective description of an island model necessarily inherits the
micro-causality of its bulk dual.

The purpose of the criterion is not to introduce new causal principles.
Instead, it is to make explicit a small set of structural assumptions that are
often left implicit, but that must be simultaneously satisfied if effective
spacelike separation is to correctly encode the absence of any physical causal
influence.
When these assumptions hold, effective micro-causality follows directly from
bulk micro-causality.

Throughout this section we restrict attention to a semiclassical regime in which
the bulk admits a local effective field theory description with a well-defined
light-cone structure.
We also assume that bulk reconstruction is meaningful within a suitable code
subspace.
Situations in which quantum gravitational effects invalidate a local notion of
bulk causality lie outside the scope of the present discussion.

\vspace{0.6em}
\noindent
\textbf{Setup.}
We consider an operator $\mathcal O_Q(p)$ associated with a codimension-one
surface $Q$, and an operator $\widetilde{\mathcal O}_\Sigma(p')$ localized on the
asymptotic boundary $\Sigma$.
The surface $Q$ may represent a Planck brane, an end-of-the-world brane, or a
genuine defect worldvolume.
An observer confined to the effective boundary description does not have access
to the bulk radial direction and therefore characterizes spacetime separation
using an intrinsic notion $\Delta_{\rm eff}(p,p')$.
Effective micro-causality requires that
\begin{equation}
\Delta_{\rm eff}(p,p')>0
\qquad \Longrightarrow \qquad
\big[\mathcal O_Q(p),\widetilde{\mathcal O}_\Sigma(p')\big]=0 ,
\label{eq:microcausality-goal}
\end{equation}
namely that operators commute whenever they are declared spacelike separated by
the effective geometry.

We assume that the effective description admits a semiclassical bulk dual
governed by a local quantum field theory for a bulk field $\phi(X)$.
Bulk locality implies the standard micro-causality condition
\begin{equation}
\big[\phi(X),\phi(X')\big]=0
\qquad
\text{whenever }
\Delta_{\rm bulk}(X,X')>0 ,
\label{eq:bulk-micro}
\end{equation}
where $\Delta_{\rm bulk}(X,X')$ denotes the invariant bulk spacetime separation.
This bulk light cone is the only fundamental causal structure in the theory.
Any notion of causality in an effective description must ultimately be inherited
from it.

\vspace{0.8em}
\noindent
\textbf{(L) Localization: no new propagation channels.}
The first ingredient is purely dynamical.
We require that degrees of freedom associated with the surface $Q$ do not
introduce new ways for signals to propagate through the bulk.
Excitations created on $Q$ may propagate along the worldvolume of $Q$ and may
interact locally with bulk fields, but they must not give rise to additional bulk
light cones.
Equivalently, there must be no physical propagation channel by which an
excitation created on $Q$ can influence $\Sigma$ except through the propagation
already allowed by the local bulk dynamics, together with interactions localized
on $Q$.

\vspace{0.8em}
\noindent
\textbf{(D) Dictionary: bulk-supported operator definition.}
The second ingredient concerns how effective operators are related to bulk
degrees of freedom.
For the causality analysis, we do \emph{not} assume a full equivalence between
effective operator algebras and bulk operator algebras.
In particular, we do not assume that operators associated with $Q$ form an
independent algebra whose commutators are defined without reference to the bulk.

Instead, we assume only the following minimal dictionary condition.
For each effective operator insertion $p\in Q$, there exists a local bulk
operator supported at a finite bulk point $X_Q(p)$ such that, for any state
$|\psi\rangle$ in the semiclassical code subspace,
\begin{equation}
\mathcal O_Q(p)\,|\psi\rangle
\;=\;
\mathcal O_{\rm bulk}\!\left(X_Q(p)\right)\,|\psi\rangle .
\label{eq:D}
\end{equation}
This relation is understood only as an equality of action on the code subspace.
It does \emph{not} represent an operator identity on the full Hilbert space.
Its content is simply that operators associated with $Q$ can be represented, in
the semiclassical regime, by inserting an appropriate local bulk operator at a
finite bulk location.
This is the minimal input needed to reduce effective commutators involving
$\mathcal O_Q$ to commutators of local bulk fields, to which bulk micro-causality
applies.
Operators on the asymptotic boundary $\Sigma$ are defined by the standard
extrapolate dictionary.

\vspace{0.8em}
\noindent
\textbf{(M) Matching: faithful interpretation of spacetime separation.}
The third ingredient concerns how spacetime separation is interpreted in the
effective description.
Condition \textbf{(M)} is not a causal assumption, but a consistency requirement.
Once the allowed propagation channels are fixed by \textbf{(L)}, and effective
operators are anchored to bulk insertions by \textbf{(D)}, effective spacelike
separation must exclude all bulk causal curves that are physically realizable.
Concretely, whenever $\Delta_{\rm eff}(p,p')>0$, there must exist no dynamically
accessible bulk causal curve connecting the corresponding bulk points
$X_Q(p)$ and $X_\Sigma(p')$.

\vspace{1.0em}
\noindent
\textbf{Sufficiency of the LDM criterion.}
When conditions \textbf{(L)}, \textbf{(D)}, and \textbf{(M)} are simultaneously
satisfied, effective micro-causality follows immediately.
Condition \textbf{(D)} allows effective commutators to be reduced to commutators
of local bulk fields.
Condition \textbf{(L)} ensures that no additional propagation channels contribute.
Condition \textbf{(M)} guarantees that effective spacelike separation excludes all
bulk causal curves that could physically mediate an influence.
Bulk micro-causality \eqref{eq:bulk-micro} then implies the vanishing of the
effective commutator \eqref{eq:microcausality-goal}.

\vspace{1.0em}
\noindent
\textbf{Bulk-first double holography.}
In bulk-first double holography, operators associated with the Planck brane admit
bulk-supported representatives at finite bulk locations, and the bulk dynamics
remains local and causal.
No independent propagation channels beyond those of the bulk theory are
introduced.
As a result, conditions \textbf{(L)} and \textbf{(D)} are satisfied.
The apparent violation of effective micro-causality arises solely because the
effective assignment of spacetime separation fails to exclude bulk causal curves
that are physically accessible.
That is, condition \textbf{(M)} fails.

\vspace{0.8em}
\noindent
\textbf{Boundary-first double holography.}
In boundary-first double holography, by contrast, the Planck brane $Q$ is treated
as an independent asymptotic boundary equipped with its own intrinsic boundary
operator algebra.
Operators on $Q$ are defined autonomously, rather than as bulk-supported
insertions at finite bulk locations.
As a consequence, condition \textbf{(D)} fails already at the operator level.
Effective commutators involving $\mathcal O_Q$ are no longer controlled by bulk
locality, even within a semiclassical regime.

Once condition \textbf{(D)} is violated, effective micro-causality is no longer
expected to follow from bulk micro-causality.
In this case, the effective description contains additional operator data and
causal structure not fixed by the bulk dynamics.
Boundary-first double holography therefore lies outside the regime of
applicability of the LDM criterion and does not constitute a counterexample to it.

%%%%%%%%%%%%%%%%%%%%%%%%%%%%%%%%%%%%%%%%%%%%%%%%%%%%%%%%%%%%%%%%%%%%%%%%%%%%%%
%%%%%%%%%%%%%%%%%%%%%%%%%%%%%%%%%%%%%%%%%%%%%%%%%%%%%%%%%%%%%%%%%%%%%%%%%%%%%%
\section{Causality in brane world constructions}
%%%%%%%%%%%%%%%%%%%%%%%%%%%%%%%%%%%%%%%%%%%%%%%%%%%%%%%%%%%%%%%%%%%%%%%%%%%%%%

We now apply the causality criterion formulated in
Section~\ref{subsec:causality-criterion} to a class of island realizations that
are conceptually distinct from double holography, namely \emph{brane world
constructions}.
By this we mean setups in which quantum field-theoretic degrees of freedom are
dynamically confined to a codimension-one hypersurface embedded in a
higher-dimensional bulk spacetime, while the bulk itself is governed by a local
and causal gravitational theory.

This class includes, as a concrete representative example, the
defect-extremal-surface (DES) construction~\cite{Deng:2020ent}, but the logic of
the argument does not rely on any special feature of that particular model.
Rather, it applies to any brane world realization in which the brane is treated
as a genuine defect in the bulk, rather than as an independent asymptotic
boundary endowed with its own autonomous boundary theory.

\vspace{0.6em}
\noindent
\textbf{Localization and the absence of bulk propagation from the brane.}
A defining structural property of brane world constructions is the strict
localization of field-theoretic degrees of freedom on the brane.
Excitations created on the brane do not propagate through the bulk interior as
independent degrees of freedom.
Instead, they propagate only along the brane worldvolume and interact with bulk
fields solely through local couplings or boundary conditions imposed at the
brane.

As a consequence, the brane does not introduce any additional bulk light cone
beyond those already present in the ambient gravitational theory.
There exists no physical propagation channel by which an excitation created on
the brane can travel through the bulk interior and reach the asymptotic boundary.
This property directly implements condition \textbf{(L)} of the causality
criterion.

\vspace{0.6em}
\noindent
\textbf{Geometric consequence: non-intersection of causal wedges.}
The DES construction provides a particularly transparent realization of this
structure.
In DES, the end-of-the-world (EOW) brane is treated as a genuine defect rather
than as an additional boundary.
Although the brane supports quantum field-theoretic degrees of freedom, these
degrees of freedom do not define independent bulk causal trajectories.

Because of this localization, any bulk causal curve that originates on the
brane and reaches the asymptotic boundary must follow propagation channels
already present in the bulk theory together with interactions localized on the
brane.
No additional shortcuts through the bulk interior are available.

\begin{figure}[h]
  \centering
  \includegraphics[width=12cm,height=8cm]{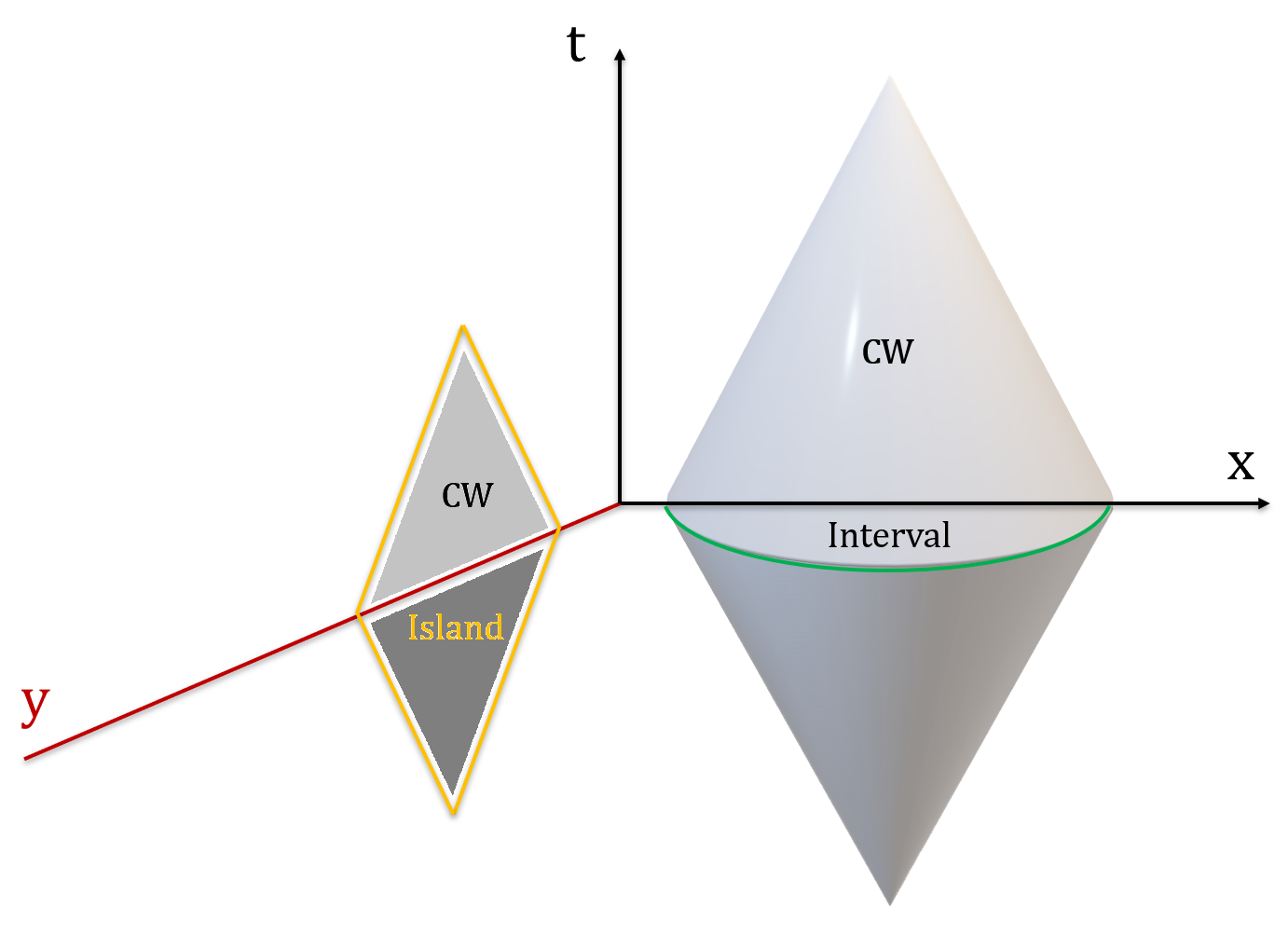}
  \caption{In brane world constructions with localized field-theory degrees of
freedom, the causal wedge of an island region on the brane can never intersect
the causal wedge of a bath interval on the asymptotic boundary.
This geometric property follows directly from the absence of independent bulk
propagation channels from the brane to the bath.}
  \label{ccw}
\end{figure}

As a result, the causal wedge associated with an island region on the brane can
never intersect the causal wedge of a bath interval on the asymptotic boundary.
This remains true even when the brane is tilted relative to $\Sigma$.
The non-intersection of causal wedges is therefore not an assumption, but a
direct geometric consequence of localization, as illustrated in
Fig.~\ref{ccw}.

\vspace{0.6em}
\noindent
\textbf{Matching of effective and bulk causal structure.}
This geometric property has an immediate implication for micro-causality.
Since no dynamically accessible bulk causal curve connects a point on the brane
to a point on the asymptotic boundary outside the effective light cone, the bulk
invariant separation relevant for commutators coincides with the intrinsic
separation assigned in the effective description.
In particular, for mixed insertions on $Q$ and $\Sigma$, the bulk invariant
separation reduces to
\begin{equation}
\widetilde{\Delta \beta}(p,p')
=
-(t-t')^2+(x-y')^2 ,
\end{equation}
which is identical to the effective separation $\Delta \beta(p,p')$.
The matching condition \textbf{(M)} is therefore automatically satisfied: once
localization removes independent propagation channels, effective spacelike
separation faithfully encodes the absence of physically realizable bulk causal
curves.

\vspace{0.6em}
\noindent
\textbf{Operator dictionary and bulk support.}
The operator-level dictionary condition \textbf{(D)} is also naturally satisfied
in brane world constructions.
Operators on the asymptotic boundary are defined by the standard extrapolate
dictionary and correspond to independent boundary degrees of freedom.
Operators localized on the brane, by contrast, are not defined by promoting the
brane to a second asymptotic boundary with its own intrinsic operator algebra.

Instead, brane operators admit bulk-supported representations with definite
geometric support on the brane worldvolume.
Although the associated degrees of freedom are dynamically confined to propagate
along the brane, the operators themselves are defined through local couplings to
bulk fields evaluated at finite bulk points $X_Q(p)$ on the brane.
Localization of degrees of freedom therefore constrains the available propagation
channels, but does not obstruct the existence of a well-defined bulk operator
embedding.

Nonlocal reconstructions of brane operators in terms of bath degrees of freedom
may exist within a suitable code subspace.
Such reconstructions reflect the quantum error-correcting structure of
holography and do not correspond to physical signal propagation.
In particular, they do not introduce independent bulk degrees of freedom or new
causal channels.
Conditions \textbf{(L)} and \textbf{(D)} therefore remain intact.

\vspace{0.8em}
\noindent
\textbf{Causal consistency of brane world islands.}
Taken together, brane world constructions with genuinely localized
field-theoretic degrees of freedom satisfy all three components of the causality
criterion:
localization of dynamics on the brane \textbf{(L)}, a bulk-supported operator
dictionary \textbf{(D)}, and automatic matching between effective and bulk
notions of spacetime separation \textbf{(M)}.
As a result, effective micro-causality is guaranteed.

This should be contrasted with double holography.
In bulk-first double holography, effective non-causality arises from a failure of
the matching condition \textbf{(M)}, despite the existence of bulk-supported
operators.
In boundary-first double holography, the dictionary condition \textbf{(D)} fails
already at a more fundamental level.
Brane world constructions such as DES avoid both pitfalls by treating the brane
as a defect with localized dynamics rather than as an independent boundary.

%%%%%%%%%%%%%%%%%%%%%%%%%%%%%%%%%%%%%%%%%%%%%%%%%%%%%%%%%%%%%%%%%%%%%%%%%%%%%%
\section{On the necessity of the causality criterion}
\label{sec:necessity}
%%%%%%%%%%%%%%%%%%%%%%%%%%%%%%%%%%%%%%%%%%%%%%%%%%%%%%%%%%%%%%%%%%%%%%%%%%%%%%

Having established in the previous sections that conditions
\textbf{(L)}, \textbf{(D)}, and \textbf{(M)} are sufficient to guarantee
micro-causality in effective descriptions of island models, we now address the
complementary question of structural necessity.
Specifically, we ask whether violations of effective micro-causality can arise
only if at least one of these structural conditions fails.
We argue that this is indeed the case.

The key point is that the three conditions constrain logically distinct and
exhaustive layers at which causal consistency may break down.
Condition \textbf{(L)} constrains the \emph{dynamical layer}, namely the
availability of physical propagation channels.
Condition \textbf{(D)} constrains the \emph{operator-theoretic layer}, namely the
existence of a bulk-supported representation of effective operators.
Condition \textbf{(M)} constrains the \emph{interpretational layer}, namely the
assignment of spacetime separation once the dynamics and operator support are
fixed.
Any violation of effective micro-causality must therefore originate from a
failure at one (or more) of these three layers.

\vspace{0.6em}
\noindent
\textbf{Failure of (L): emergence of independent propagation channels.}
Suppose that condition \textbf{(L)} is violated, so that operators associated with
the surface $Q$ generate autonomous excitations capable of propagating through
the bulk interior as independent degrees of freedom.
In this case, the brane or defect supports additional causal channels that define
bulk light cones not present in the ambient gravitational theory.
Excitations created on $Q$ may then reach the asymptotic boundary through
propagation paths that are not encoded in the effective description.

From the perspective of an observer confined to the effective boundary geometry,
such channels are invisible.
Two operator insertions may therefore be declared spacelike separated according
to $\Delta_{\rm eff}$, while being timelike related through the additional bulk
propagation channel.
The corresponding commutator does not vanish, and effective micro-causality is
violated.
This failure mechanism is purely dynamical and does not rely on any ambiguity in
the operator dictionary or in the interpretation of spacetime separation.

\vspace{0.6em}
\noindent
\textbf{Failure of (D): loss of local bulk operator support.}
Next, consider a violation of condition \textbf{(D)}, which is precisely what
occurs in \emph{boundary-first double holography}.
In this formulation, operators associated with the surface $Q$ are not defined
as bulk-supported insertions at finite bulk locations.
Instead, $Q$ is treated as an independent asymptotic boundary, and operators on
$Q$ are endowed with their own intrinsic boundary operator algebra, on the same
footing as operators on $\Sigma$.

Under these circumstances, commutators of effective operators involving
$\mathcal O_Q$ cannot, in general, be reduced to commutators of local bulk fields
with well-defined geometric support.
As a result, bulk locality no longer controls the effective operator algebra,
even if the bulk dynamics itself remains local and causal.
Violations of effective micro-causality may then arise without invoking any
additional propagation channels.

In boundary-first double holography, this is not a dynamical pathology but a
structural one.
Once the operator-level anchoring to local bulk degrees of freedom is abandoned,
effective causality is no longer inherited from bulk micro-causality.
Whether the effective description is causal must instead be determined
intrinsically within the boundary theory itself, as the bulk no longer provides
a priori control over operator commutators.
\paragraph{Bulk duality versus causal inheritance.}
It is important to emphasize that the existence of a bulk dual description is
not, by itself, sufficient to guarantee that an effective boundary theory
inherits the micro-causality of the bulk.
Bulk micro-causality constrains the commutators of \emph{local bulk operators}
anchored at definite spacetime points.
To translate this constraint into a statement about effective operator algebras,
one must be able to associate effective operators with bulk-supported
representatives localized at finite bulk loci.

This distinction is automatic for asymptotic boundaries, where the extrapolate
dictionary provides a canonical identification between boundary operator
insertions and bulk fields approaching the boundary along fixed directions.
By contrast, in boundary-first double holography the Planck brane is treated as
an autonomous boundary endowed with its own intrinsic operator algebra.
Although the brane theory admits a bulk gravitational dual at the level of states
and observables, individual brane operators are not required to be anchored to
local bulk insertion points.
As a result, bulk micro-causality does not, in general, constrain the effective
operator algebra on the brane.

From this perspective, boundary-first double holography does not represent a
failure of bulk causality, but rather lies outside the regime in which bulk
causal structure can be cleanly inherited by an effective description.

\vspace{0.6em}
\noindent
\textbf{Failure of (M): mismatch between effective and bulk spacetime separation.}
Finally, suppose that conditions \textbf{(L)} and \textbf{(D)} are satisfied, but
condition \textbf{(M)} is violated.
In this case, the effective description assigns spacelike separation to pairs of
points $(p,p')$ for which there exists a dynamically accessible bulk causal curve.
Although no independent propagation channels are present and effective operators
admit bulk-supported representations, the effective notion of spacetime
separation fails to faithfully encode bulk causal accessibility.

This mechanism underlies the non-causality of bulk-first double holography.
As shown in Section~\ref{sec:noncausality-dh}, operators localized on the Planck
brane and on the asymptotic boundary may be declared spacelike separated according
to the effective boundary geometry, while being timelike related in the bulk.
The resulting non-vanishing commutator reflects a breakdown of condition
\textbf{(M)}, rather than any failure of bulk locality or of the operator
dictionary.
By contrast, formulations in which the Planck brane is treated as an independent
boundary fall into the failure mode of condition \textbf{(D)}, as discussed
earlier.

\vspace{0.6em}
Taken together, these three failure modes exhaust all logically distinct ways in
which effective micro-causality can break down.
Any violation must originate from either the presence of independent propagation
channels, the loss of local bulk operator support, or an inconsistent assignment
of spacetime separation.
Conversely, if conditions \textbf{(L)}, \textbf{(D)}, and \textbf{(M)} are all
satisfied, none of these mechanisms is available.
It follows that the LDM criterion is not only sufficient but also necessary for
an effective island description to faithfully inherit the causal structure of
its bulk dual.
The criterion therefore provides a complete structural characterization of
micro-causality in island models.

%%%%%%%%%%%%%%%%%%%%%%%%%%%%%%%%%%%%%%%%%%%%%%%%%%%%%%%%%%%%%%%%%%%%%%%%%%%%%%
%%%%%%%%%%%%%%%%%%%%%%%%%%%%%%%%%%%%%%%%%%%%%%%%%%%%%%%%%%%%%%%%%%%%%%%%%%%%%%
\section{Causality in time-dependent island configurations}
\label{sec:time-dependent}
%%%%%%%%%%%%%%%%%%%%%%%%%%%%%%%%%%%%%%%%%%%%%%%%%%%%%%%%%%%%%%%%%%%%%%%%%%%%%%

In this section we examine the stability of the causality criterion under
time-dependent evolution.
Throughout the discussion we restrict attention to semiclassical backgrounds in
which the bulk admits a well-defined local causal structure at each instant of
time.
Possible state dependence of operator reconstruction plays no role in the
analysis, since the criterion is formulated in terms of causal accessibility
rather than operator identities.

The analysis so far has focused primarily on static or time-independent
configurations.
However, island models of physical interest are intrinsically dynamical: islands
may appear, disappear, or move as the state evolves, for instance during black
hole evaporation.
It is therefore essential to ask whether time dependence introduces qualitatively
new mechanisms by which effective micro-causality could fail.

The central claim of this section is that it does not.
Time dependence by itself does not introduce any new obstruction to effective
micro-causality beyond those already identified in the static setting.
When the structural conditions \textbf{(L)}, \textbf{(D)}, and \textbf{(M)} are
maintained throughout the evolution, effective micro-causality is preserved at
all times.
Conversely, any apparent time-dependent non-causality can always be traced to a
violation of one of these conditions, rather than to time dependence per se.

\vspace{0.4em}
In a time-dependent setting, both the bulk geometry and the location of quantum
extremal surfaces evolve.
As a result, the island associated with a given bath region may change over time.
Nevertheless, at each instant the bulk dynamics is governed by a local and causal
field theory.
In particular, the bulk micro-causality condition \eqref{eq:bulk-micro} continues
to hold pointwise in spacetime.
Any potential violation of effective micro-causality must therefore originate
from a failure of one of the structural conditions identified previously, rather
than from the presence of time dependence itself.

\vspace{0.6em}
\noindent
\textbf{Localization under time evolution.}
We begin with condition \textbf{(L)}.
In brane world constructions with localized field-theory degrees of freedom, time
dependence does not introduce new autonomous propagation channels.
Excitations created on the brane continue to propagate only along the brane
worldvolume, even when the embedding of the brane evolves in time.
No additional bulk light cone is generated dynamically.
Accordingly, time dependence alone cannot lead to a violation of condition
\textbf{(L)}.

It is useful to contrast this with bulk-first double holography.
There, operators associated with the island are bulk-supported rather than
defect-localized, and the notion of localization in \textbf{(L)} is therefore not
violated but instead inapplicable.
Time dependence does not change this structural distinction: whether or not
independent propagation channels exist is determined by how degrees of freedom
are defined, not by whether the configuration is static or dynamical.

\vspace{0.6em}
\noindent
\textbf{Dictionary under time evolution.}
Next, consider condition \textbf{(D)}.
The operator-level embedding of effective degrees of freedom into the bulk is
defined locally in spacetime.
In time-dependent backgrounds, operators localized on the brane or on the
asymptotic boundary continue to admit bulk-supported representatives at their
respective spacetime points.
Although the explicit form of the dictionary may evolve with time, this evolution
does not alter the basic fact that effective operators are supported on finite
bulk loci.

As in the static case, nonlocal reconstruction of operators within a code
subspace reflects the quantum error-correcting structure of holography and does
not introduce new dynamical degrees of freedom or new causal channels.
Consequently, condition \textbf{(D)} is stable under time evolution in both
brane world constructions and bulk-first double holography.
By contrast, in boundary-first double holography, condition \textbf{(D)} fails
already at the kinematical level, and time dependence does not ameliorate this
failure.

\vspace{0.6em}
\noindent
\textbf{Matching under time evolution.}
The most delicate issue concerns condition \textbf{(M)}.
In time-dependent geometries, the effective notion of spacelike separation
evolves as the island configuration changes.
Effective spacelike separation must therefore be defined with respect to the
dynamically available propagation channels at the time of operator insertion.

Provided this assignment is made locally in time and consistently with the bulk
geometry, effective spacelike separation continues to exclude all physically
realizable bulk causal curves.
In this case, the matching condition \textbf{(M)} remains satisfied, and effective
micro-causality is preserved.

Apparent paradoxes can arise if effective spacetime separation is defined using a
fixed-time or quasi-static geometry while bulk propagation is implicitly allowed
according to a different stage of the evolution.
Such inconsistencies do not signal a genuine breakdown of micro-causality, but
rather an inconsistent comparison between different time slices.
When the effective causal structure is defined covariantly and locally in time,
the matching condition \textbf{(M)} remains intact.

\vspace{0.6em}
Taken together, these observations imply that time dependence does not introduce
any new failure mode beyond those already classified in
Section~\ref{sec:necessity}.
Violations of effective micro-causality can occur only if one of the structural
conditions \textbf{(L)}, \textbf{(D)}, or \textbf{(M)} is dynamically violated
during the evolution.
In particular, island formation and evaporation reflect changes in the
entanglement structure of the state rather than a breakdown of causal
consistency.

We therefore conclude that the causality criterion formulated in
Section~\ref{subsec:causality-criterion} is robust under time evolution.
When the underlying bulk dynamics remains local and the effective description
respects localization, dictionary, and matching at each instant of time,
micro-causality is preserved throughout time-dependent island processes.
This establishes the causal consistency of dynamical island models realized as
brane world constructions and clarifies the origin of apparent time-dependent
paradoxes in bulk-first double holography.

\section{Conclusion and discussion}
\label{sec:conclusion}
%%%%%%%%%%%%%%%%%%%%%%%%%%%%%%%%%%%%%%%%%%%%%%%%%%%%%%%%%%%%%%%%%%%%%%%%%%%%%%

In this work we have undertaken a systematic analysis of micro-causality in island
models, with the goal of isolating the precise structural conditions under which
an effective description faithfully inherits the causal consistency of its bulk
dual.
Our central conclusion is that apparent violations of micro-causality are neither
an intrinsic feature of island physics nor a consequence of nonlocal entanglement
or operator reconstruction.
Instead, they arise from a structural mismatch between effective and bulk notions
of spacetime separation when certain basic requirements are not satisfied.

A key step in our analysis was to disentangle two conceptually distinct notions of
double holography that are often conflated in the literature.
In what we termed \emph{bulk-first} double holography, operators associated with
the island are fundamentally bulk-supported, and the effective boundary
description inherits its dynamics from the bulk.
In this setting, the bulk theory remains fully local and causal, and effective
operators admit well-defined bulk representatives.
The apparent violation of micro-causality arises solely because the effective
assignment of spacetime separation fails to exclude dynamically accessible bulk
causal curves.
By contrast, in \emph{boundary-first} double holography the Planck brane is treated
as an independent asymptotic boundary endowed with its own autonomous operator
algebra.
In this formulation, the minimal operator-level control required to inherit bulk
micro-causality is lost from the outset.
Distinguishing these two structures is essential for a coherent discussion of
causality in island models.

We further clarified a common source of confusion by sharply separating nonlocal
operator reconstruction from causal signal propagation.
Reconstruction of island or brane operators from bath degrees of freedom reflects
the quantum error-correcting structure of holography within a restricted code
subspace.
It does not introduce new physical propagation channels and therefore poses no
threat to micro-causality.
Apparent tensions arise only if reconstruction is implicitly conflated with
dynamical signal transmission.

Motivated by these observations, we formulated a structural criterion for
micro-causality in effective descriptions of island models.
The criterion consists of three logically distinct components:
\textbf{(L)} the absence of independent propagation channels beyond those present
in the bulk theory,
\textbf{(D)} the existence of a bulk-supported operator definition for effective
degrees of freedom,
and \textbf{(M)} a consistent matching between effective spacelike separation and
dynamically accessible bulk causal curves.
When these conditions are satisfied, effective micro-causality follows directly
from bulk micro-causality.
Conversely, we showed that violations of effective causality can arise only if at
least one of these conditions fails.
In this sense, the LDM criterion provides both a sufficient and a necessary
structural characterization of micro-causality in island models.

We applied this criterion to brane world constructions, with the
defect-extremal-surface construction serving as a representative example.
In such models, quantum field theory degrees of freedom are genuinely localized on
a codimension-one hypersurface and do not define autonomous bulk light cones.
As a result, effective and bulk notions of spacetime separation coincide once the
correct dynamical interpretation is adopted, and micro-causality is preserved.
The contrast with bulk-first double holography is therefore structural rather than
model-dependent, while boundary-first double holography lies outside the regime
of applicability of the criterion altogether.

It is worth emphasizing that this perspective is complementary to recent
top-down analyses of causality in double holography.
In fully controlled string-theoretic constructions, apparent causality puzzles
are resolved by carefully distinguishing the holographic dual of the full BCFT
from the dual obtained by dualizing only defect degrees of freedom, and by
recognizing that the latter is not realized as a simple subsector or geometric
subregion of the former \cite{Karch:2022rvr}.
From that viewpoint, the paradox arises from an overly naive identification of the
``brane theory'' with part of the full BCFT dual geometry.
By contrast, our analysis does not rely on refining the holographic dictionary
itself, but instead focuses on the causal interpretation of effective
descriptions.
We argue that, once effective spacelike separation is correctly matched to
dynamically accessible bulk causal curves, no genuine violation of
micro-causality arises even at the level of the brane or defect-extremal-surface
effective theory.
In this sense, the two approaches address different aspects of the same apparent
tension and lead to mutually consistent conclusions.

We also examined the role of time dependence, which is essential for applications
to black hole evaporation and information transfer.
We found that time dependence does not introduce new mechanisms for violating
micro-causality beyond those already present in static configurations.
Provided that localization, dictionary, and matching are maintained locally in
time, island formation, motion, and disappearance remain compatible with causal
consistency throughout the evolution.

More broadly, our results highlight a general lesson for emergent spacetime in
holographic systems.
Causal structure in an effective description is not guaranteed by entanglement
structure or operator reconstruction alone.
It emerges only when effective notions of locality and separation are properly
aligned with the dynamical causal structure of the underlying bulk theory.
When this alignment fails, effective non-causality can arise even in the presence
of a perfectly local and causal bulk description.

There are several natural directions for future investigation.
It would be interesting to apply the LDM criterion to higher-dimensional island
models, to setups involving multiple defects or branes, and to scenarios with more
general boundary conditions.
Another important direction is to understand how the criterion should be modified
in regimes where quantum gravitational effects blur the notion of a sharp bulk
light cone.
More generally, the structural perspective developed here may prove useful
whenever effective spacetime notions are employed in emergent descriptions of
quantum gravity, well beyond the specific context of island models.

\section*{Acknowledgments}
Feiyu Deng is  supported by the National Natural Science Foundation of China (NSFC) Project No.12547135.

\appendix

\bibliographystyle{JHEP}
\bibliography{islandttbar}

\end{document}